\documentclass[aps,prl,12pt]{revtex4-1}

\usepackage{graphicx,graphics,color,epsfig}
\usepackage{amsmath}
\usepackage{amssymb}
\usepackage{amsfonts}
\usepackage{amsfonts}
\usepackage{epstopdf}
\usepackage{bm}
\usepackage{times,xspace}
\usepackage{color}

\begin{document}
\title{Spin-orbit density wave induced hidden topological order in URu$_2$Si$_2$}
\author{Tanmoy Das$^{1,2}$\\
$^1$Theoretical Division, Los Alamos National Laboratory, Los Alamos, NM 87545, USA.\\
$^2$ Physics Department, Northeastern University, Boston, MA 02115, USA.}
\begin{abstract}
\textbf{The conventional order parameters in quantum matters are often characterized by `spontaneous' broken symmetries. However, sometimes the broken symmetries may blend with the invariant symmetries to lead to mysterious emergent phases. The heavy fermion metal URu$_2$Si$_2$ is one such example, where the order parameter responsible for a second-order phase transition at $T_h$=17.5~K has remained a long-standing mystery. Here we propose via {\it ab-initio} calculation and effective model that a novel spin-orbit density wave in the $f$-states is responsible for the hidden-order phase in URu$_2$Si$_2$. The staggered spin-orbit order `spontaneous' breaks rotational, and translational symmetries while time-reversal symmetry remains intact. Thus it is immune to pressure, but can be destroyed by magnetic field even at $T=0$~K, that means at a quantum critical point. We compute topological index of the order parameter to show that the hidden order is topologically invariant. Finally, some verifiable predictions are presented.
}
\end{abstract}
\maketitle \narrowtext

Most states or phases of matter can be described by local order parameters and the associated broken symmetries in the spin, charge, orbital or momentum channel. However, recent discoveries of quantum Hall states,\cite{SCZSHE} and topological insulators\cite{FuKane,SCZBiSe} have revamped this conventional view. It has been realized\cite{SCZSHE,FuKane,SCZBiSe,TI_exp} that systems with combined time-reversal ($\mathcal{TR}$) symmetry and large spin-orbit (SO) coupling can host new states of matter which are distinguished by topological quantum numbers of the bulk band structure rather than spontaneously broken symmetries. Subsequently, more such distinct phases have been proposed in the family of topological Mott insulators,\cite{TI_Mott} topological Kondo insulators,\cite{TI_Kondo} topological antiferromagnetic insulators,\cite{TI_AFM}. In the latter cases, the combined many-body physics and $\mathcal{TR}$ symmetry governs topologically protected quantum phases. Encouraged by these breakthrough developments, we search for analogous exotic phases in the heavy fermion metal URu$_2$Si$_2$, whose low-energy $f$ states accommodate ${\mathcal{TR}}$ and strong SO coupling. This compound also naturally hosts diverse quantum mechanical phases including Kondo physics, large moment antiferromagnetism (LMAF), mysterious `hidden-order' (HO) state, and superconductivity.\cite{Mydosh_review}.

In URu$_2$Si$_2$ the screening of $f$-electrons due to the Kondo effect begins at relatively high temperatures, ushering the system into a heavy fermion metal at low-temperature.\cite{STM_Yazdani} Below $T_h$=17.5~K, it enters into the HO state via a second-order phase transition characterized by sharp discontinuities in numerous bulk properties.\cite{Cv_Palstra,rho_Palstra,nonlin_sus2,thermal} The accompanying gap is opened both in the electronic structure\cite{STM_Yazdani,STM_Davis,ARPES,ARPES_FS} as well as in the magnetic excitation spectrum,\cite{MasonINS} suggesting the formation of an itinerant magnetic order at this temperature. However, the associated tiny moment ($\sim0.03\mu_B$) cannot account for the large (about 24\%) entropy release\cite{Broholm} and other sharp thermodynamic\cite{Cv_Palstra,rho_Palstra} and transport anomalies\cite{nonlin_sus2,thermal} during the transition. Furthermore, very different evolutions of the HO parameter and the magnetic moment as a function of both magnetic field\cite{Mentink,ChandraColeman} and pressure\cite{Amitsuka,Fisher} rule out a possible magnetic origin of the HO phase in this system. Any compelling evidence for other charge, orbital or structural ordering has also not been obtained.\cite{Matsuda} Existing theories include multiple spin correlator\cite{Barzykin}, Jahn-Teller distortions\cite{Kasuya}, unconventional spin-density wave,\cite{Ikeda,Mineev}, antiferromagnetic fluctuation,\cite{Okuno} orbital order, \cite{ChandraColeman}, helicity order,\cite{Varma}, staggered quadrupole moment\cite{Santini}, octupolar moment,\cite{Hanzawa} hexadecapolar order,\cite{Kotliar}, linear antiferromagnetic order,\cite{Oppener} incommensurate hybridization wave,\cite{Sasha} spin nematic order,\cite{SN} modulated spin liquid,\cite{Pepin} $j$-$j$ fluctuations,\cite{Oppeneer2}, unscreened Anderson lattice model,\cite{Riseborough}  among others\cite{Mydosh_review}. However, a general consensus for the microscopic origin of the HO parameter has not yet been attained.

Formulating the correct model for the HO state requires the knowledge of the broken symmetries and the associated electronic degrees of freedom that are active during this transition. A recent torque measurement on high quality single crystal sample reveals that the four-fold rotational symmetry of the crystal becomes spontaneously broken\cite{Matsuda} at the onset of the HO state. Furthermore, several momentum-resolved spectroscopic data unambiguously indicate the presence of a translational symmetry breaking at a longitudinal incommensurate wavevector ${\bm Q}_h=(1\pm0.4,0,0)$.\cite{STM_Davis,INS,Broholm,ARPES_FS} [Previous first-principle calculation has demonstrated that an accompanying commensurate wavevector ${\bm Q}_2=(1,0,0)$ might be responsible for the LMAF phase,\cite{Oppener} which is separated from the HO state via a first order phase transition.\cite{Mydosh_review,Mentink,ChandraColeman,Amitsuka,Fisher}. As it is often unlike to have two phases of same broken symmetry but separated by a phase boundary, we expect that LMAF and HO phases are different.] In general, the order parameter that emerges due to a broken symmetry relies incipiently on the good quantum number and symmetry properties of the `parent'  or non-interacting Hamiltonian. In case of URu$_2$Si$_2$, spin and orbital are not the good quantum numbers, rather the presence of the SO coupling renders the total angular momentum to become the good quantum number. Therefore, $SU(2)$ symmetry can not be defined for spin or orbital alone, and the `parent' Hamiltonian has to be defined in $SU(2)\otimes SU(2)$ representation. The `parent' Hamiltonian also accommodate other symmetries coming from its crystal, wavefunction properties which we desire to incorporate to formulate the HO parameter.

\section{Results}

\begin{figure}[here]
\hspace{-0cm}
\rotatebox{0}{\scalebox{.8}{\includegraphics{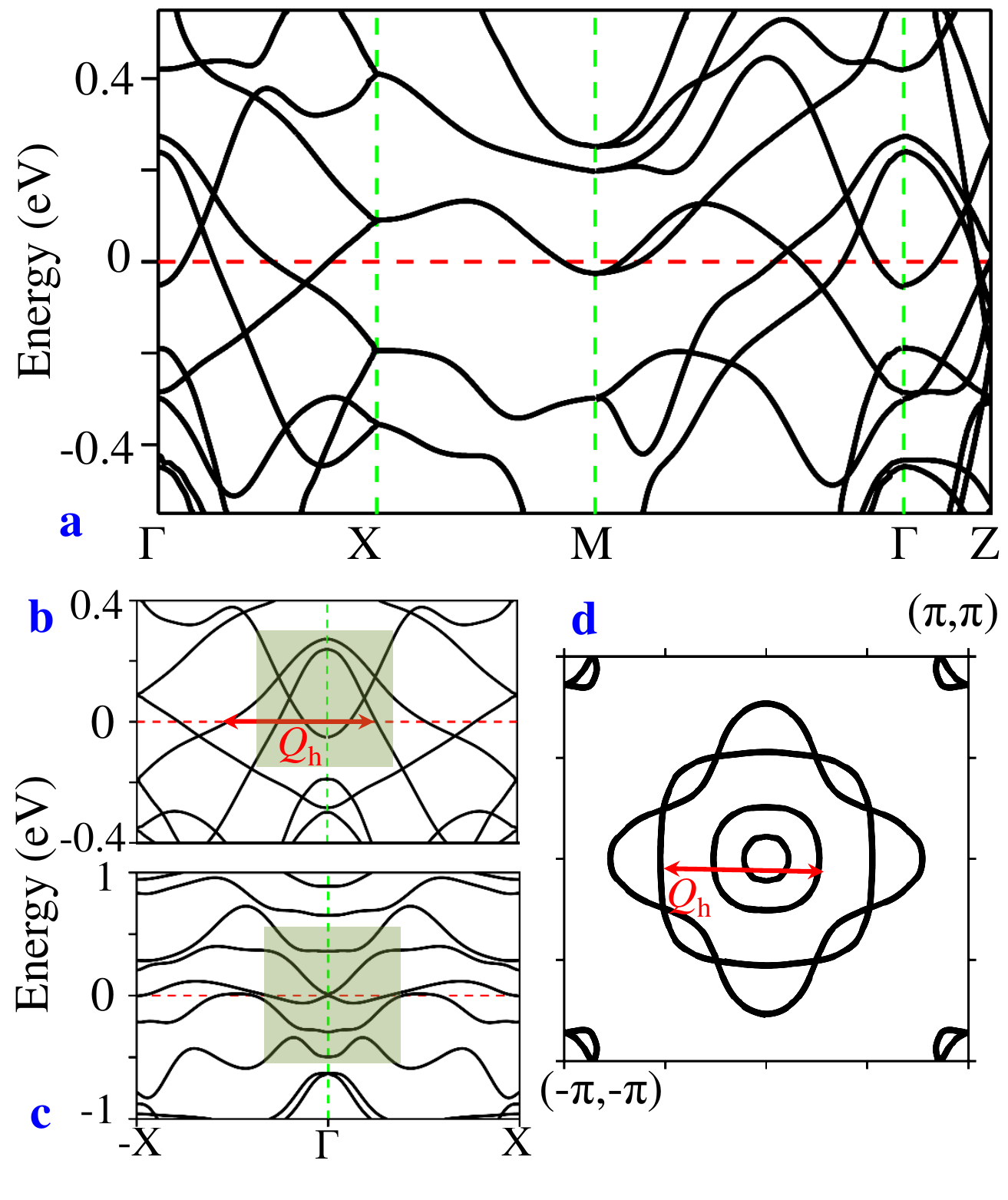}}}
\caption{{\bf {\it Ab initio} band structure and Fermi surface of URu$_2$Si$_2$.} {\bf a}, Computed non-interacting energy dispersions of URu$_2$Si$_2$, using Wien2K software,\cite{Wien,GGA} are presented along $\Gamma$(0,0,0), X($\pi$,0,0), M($\pi$,$\pi$,0), and Z(0,0,$\pi$) directions. The band structure is consistent with the previous full potential local orbitals (FPLO) and full potential linearized augmented plane wave (FPLAPW) calculations in the paramagnetic state.\cite{Oppener} The low-energy dispersions along $\Gamma$-X is expanded in {\bf b} and contrasted with the same but without the SO coupling in {\bf c}. The FS in the $k_z=0$ plane is shown in {\bf d}. The red arrow dictates the FS `hot-spot' that emerges after including SO coupling. }
\label{fig1}
\end{figure}

{\bf Ab-initio band structure.} In order to find out the symmetry properties of the low-lying states, we begin with investigating the {\em ab-initio} `parent' band dispersion and the FS of URu$_2$Si$_2$\cite{Wien,GGA} in Fig.~\ref{fig1}. The electronic structure in the vicinity of $E_F$ ($\pm$0.2eV) is dominated by the 5$f$ states of U atom in the entire Brillouin zone.\cite{Oppener,STM_Davis,ARPES,ARPES_FS,INS,Denlinger} Owing to the SO coupling and the tetragonal symmetry, the 5$f$ states split into the octet $J$=$\frac{7}{2}$ ($\Gamma_8$) states and the sextet $J$=$\frac{5}{2}$ ($\Gamma_6$) states.\cite{HottaPRB} URu$_2$Si$_2$ follows a typical band progression in which the $\Gamma_8$ bands are pushed upward to the empty states while the $\Gamma_6$ states drop to the vicinity of $E_F$. The corresponding FS in Fig.~\ref{fig1}{\bf d} reveals that an even number of anti-crossing features occurs precisely at the intersection between two oppositely dispersing conducting sheets. Unlike in topological insulators,\cite{SCZBiSe,TI_exp} the departure of the band crossing points from the $\mathcal{TR}$-invariant momenta here precludes the opening of an inverted band gap at the crossings,\cite{FuKane} and Dirac-cones crop up with Kramer's degeneracy in the bulk states. Therefore, URu$_2$Si$_2$ is an intrinsically trivial topological metal above the HO transition temperature.

The SO interaction introduces two prominent FS instabilities at $Q_2=(1,0,0)$ and at $Q_h=(1\pm0.4,0,0)$. The commensurate wavevector $Q_2$ occurs between same orbitals. Therefore, if this instability induces a gap opening, it has to be in the spin-channel, which is prohibited by $\mathcal{TR}$ symmetry and strong SO coupling. We argue (see Supplementary Information (SI) for details), in accordance with an earlier calculation,\cite{Oppener} that this instability is responsible for the LMAF phase. On the other hand, the incommensurate one, $Q_h$, occurs between two different orbitals, and can open a gap if a symmetry between these orbitals and spins are spontaneously broken together. In other word, since spin-orbit coupling is strong in this system, individual spin- or orbital-orderings are unlikely to form unless interaction can overcome the spin-orbit coupling strength. On the other hand, a SO entangled order parameter in the two-particle channel can collectively propagate with alternating sign in the total angular momentum at the wavelength determined by the modulation vector. This is the guiding instability that drives spontaneous rotational symmetry breaking, while the $\mathcal{TR}$ symmetry remains intact (see Fig.~\ref{fig2}{\bf a}). This is because, both $SU(2)$ groups for spin and orbital separately are odd under $\mathcal{TR}$, but their product $SU(2)\otimes SU(2)$ becomes even. As the parent state is not a non-trivial topological phase, a gap is opened to lift the FS instability.

{\bf Low-energy effective model.} Motivated by the above-mentioned experimental results and band structure symmetry properties, we formulate a simple and unified model by using the theory of invariants\cite{Winkler}. We restrict our discussion to the low-lying $\Gamma_6$ bands and neglect the unfilled $\Gamma_8$ bands. Due to $j$-$j$ SO coupling and $\mathcal{TR}$ symmetry, the $\Gamma_6$ atomic states consist of three doublets, characterized by up and down `pseudospins': $m_J$=$\pm$$\frac{5}{2}$,~$\pm$$\frac{3}{2}$,~$\pm$$\frac{1}{2}$, where $m_J$ is the $z$ component of $J$. On entering into the HO state, the FS instability commences in between the two doubly degenerate $|m_J|$=$\frac{3}{2}$ and $\frac{1}{2}$ states only.\cite{Oppener,SN} If no other symmetry is broken, the degenerate $|m_J|$=$\frac{5}{2}$ state remains unaltered in the HO state,\cite{Winkler} and hence they are not considered in our model Hamiltonian. Throughout this paper, we consistently use two indices: orbital index $\tau=|m_J|=\frac{1}{2} (\frac{3}{2})$, and `pseudospin' $\sigma$=$\uparrow$(+),~$\downarrow$(-). In this notation, we consider the `pseudospinor' field $\hat{\Psi}^{\dag}({\bm k})$=($f^{\dag}_{{\bm k},\frac{1}{2},+},~f^{\dag}_{{\bm k},\frac{3}{2},+},~f^{\dag}_{{\bm k},\frac{1}{2},-},~f^{\dag}_{{\bm k},\frac{3}{2},-})$, where $f^{\dag}_{\bm{k},\tau,\sigma}$ is the creation operator for an electron in the orbital $|m_J|$=$\frac{1}{2},\frac{3}{2}$ with momentum $\bm{k}$ and `pseudospin' $\sigma$.

The representation of the symmetry operations that belongs to the  $D_{4h}$ symmetry of the URu$_2$Si$_2$ crystal structure is: $\mathcal{TR}$ symmetry,  inversion symmetry $\mathcal{I}$, four-fold rotational symmetry $\mathcal{C}_4$, and the two reflection symmetries $\mathcal{P}_{x/y}$. The SO $f$-state of actinides is invariant under all symmetries except the mirror reflection, which in fact allows the formation of the SO density wave into a finite gap in the HO state (see {\it SI}). On the basis of these symmetry considerations, it is possible to deduce the general form of the non-interacting Hamiltonian as:
\begin{eqnarray}
&&H_0=\sum_{\bm{k},\sigma}\hat{\Psi}^{\dag}({\bf k})
\left(
\begin{array}{cc}\
h_{11}(\bm{k}) & h_{12}(\bm{k}) \\
h_{21}(\bm{k}) & h_{22}(\bm{k})
\end{array}
\right)
\hat{\Psi}({\bf k}),\\
&&~~~~~~h_{\tau\tau^{\prime}}(\bm{k})=\epsilon_{\tau\tau^{\prime}}({\bf k})\bm{\tau}^0 + \bm{d}_{\tau\tau^{\prime}}({\bf k})\cdot \bm{\tau}.
\end{eqnarray}
Here, $\bm{\tau}^{\mu}$ ($\mu\in0,x,y,z$) depict the 2D Pauli matrices in the orbital space  and $\tau^0$ is the identity matrix ($\bm{\sigma}^{\mu}$ matrices will be used later to define the spin space). The $\mathcal{TR}$ invariance requires that $h_{22/21}(\bm{k})$=$h^*_{11/12}(-\bm{k})$. Under  $\mathcal{TR}$ and $\mathcal{I}$, the symmetry of $\epsilon_{\tau\tau^{\prime}}({\bm k})$ and ${d}_{\tau\tau^{\prime}}^{x,y,z}({\bf k})$ must complement to their corresponding identity and Pauli Matrix counterparts, respectively. Hence we obtain the Slater-Koster hopping terms as:
$[\epsilon({\bm k}),d^x,d^y,d^z]_{11}=\big[-2t(\cos{k_x}+\cos{k_y})-\mu,~-2t_1\sin{k_x},~-2t_1\sin{k_y},~-2t_2(\cos{k_x}-\cos{k_y})\big]$, and $[\epsilon({\bf k}),d^x,d^y,d^z]_{12}=\big[0,0,0,-4t_z\cos{(k_x/2)}\cos{(k_y/2)}\cos{(k_z/2)}\big]$. The obtained values of the tight-binding hopping parameters as $(t,t_1,t_2,t_z)$=(-45,45,50,-25) in meV. The above Hamiltonian can be solved analytically which gives rise to four SO-split energy dispersions as
\begin{equation}
E^{\tau\sigma}(\bm{k})=\epsilon({\bm k})+\tau\sqrt{\sum_{\mu}|{\bm d}^{\mu}_{12}({\bf k})|^2}+\sigma\sqrt{\sum_{\mu}|{\bm d}^{\mu}_{11}({\bm k})|^2}. 
\label{eq:band}
\end{equation}
Here $\sigma=\pm$ and $\tau=\pm$ become band indices. An important difference of the present Hamiltonian with that of bulk topological insulators\cite{SCZBiSe} or quantum spin-Hall systems\cite{SCZSHE} is the absence of a mass or gap parameter in the former case. The computed non-interacting bands are plotted in Fig.~\ref{fig2}({\bf b}), which exhibit several Dirac points along the high-symmetry lines. Focusing on the Dirac point close to $E_F$, we find that it occurs at the crossing between bands $E^{+-}$ and $E^{-+}$, demonstrating that it hosts four-fold Kramer's degeneracy (two orbitals and two spins). Therefore, lifting this degeneracy requires the presence of a SO order parameter. However, it is important to note that the gap opening at the Dirac point is not a manifestation of the presence of degeneracy at it, but a consequence of the SO density wave caused by FS instability.

{\bf SO density wave induced HO.} The `hot-spot' ${\bm Q}_h$ divides the unit cell into a reduced `SO Brillouin zone' in which we can define the Nambu operator in the usual way $( \hat{\Psi}^{\dag}({\bm k})$, $\hat{\Psi}^{\dag}({\bm k}+{\bm Q}_h^x)$, $\hat{\Psi}^{\dag}({\bm k}+{\bm Q}_h^y))$. In this notation, the modulated SO density wave (SODW) term can be written in general as
\begin{eqnarray}
H_{SODW} = \sum_{\mu\nu} g^{\mu\nu} : \left[\hat{\Psi}^{\dag}(\bm{k})\Gamma^{\mu\nu}\hat{\Psi}(\bm{k}+\bm{Q}_h)\right]^2:,
\end{eqnarray}
where $\mu,\nu\in\{0,x,y,z\}$. The symbol $::$ represents normal ordering. Here $g$ is the contact coupling interaction arising from screened interorbital Coulomb term embedded in Hund's coupling parameter, and $\Gamma^{\mu\nu}={\bm \tau}^{\mu}\otimes\bm{\sigma}^{\nu}$, $\bm{\tau}$ and $\bm{\sigma}$  represent Pauli matrices in orbital and spin basis, respectively. Absorbing $g$ and $\Gamma$ into one term we define the mean-field order parameter
\begin{eqnarray}
M^{\mu\nu} &=& g^{\mu\nu}(\bm{k})\left\langle\hat{\Psi}^{\dag}({\bm k})[{\bm \tau}^{\mu}\otimes{\bm \sigma}^{\nu}]\hat{\Psi}({\bm k}+{\bm Q}_h)\right\rangle.\\
& =& g^{\mu\nu}(\bm{k})\left\langle f_{{\bm k},\tau,\sigma}^{\dag}[{\bm \tau}_{\tau\tau^{\prime}}^{\mu}{\bm \sigma}_{\sigma\sigma^{\prime}}^{\nu}]f_{{\bm k}+{\bm Q}_h,\tau^{\prime},\sigma^{\prime}}\right\rangle.
\end{eqnarray}
Here $\tau,\tau^{\prime}$ and $\sigma,\sigma^{\prime}$ (not in bold font) are the components of the ${\bm \tau}^{\mu}$ and ${\bm \sigma}^{\nu}$ matrices, respectively. Without any loss of generality we fix the spin orientation along $z$-directions ($\nu=z$).  Therefore, we drop the index $\nu$ henceforth. Furthermore we define the gap vector as
${\bm b}^{\mu}_{\tau\tau^{\prime}}({\bm k})=g^{\mu}\Delta_{\tau\tau^{\prime}}^{\mu}({\bm k}){\bm \tau}_{\tau\tau^{\prime}}^{\mu}$, where we split the interaction term $g({\bm k})$ into a constant onsite term and the dimensionless order parameter $\Delta({\bm k})$. With these substitutions, we obtain the final result for the order parameter as
\begin{eqnarray}
M^{\mu} &=& \left\langle \sum _{\tau\tau^{\prime}\sigma\sigma^{\prime}} f_{{\bm k},\tau,\sigma}^{\dag}\left[{\bm b}_{\tau\tau^{\prime}}^{\mu}({\bm k}){\bm \sigma}_{\sigma\bar{\sigma}}^z\right]f_{{\bm k}+{\bm Q}_h,\tau^{\prime},\bar{\sigma}}\right\rangle.
\label{eq:op}
\end{eqnarray}

Eq.~\ref{eq:op} admits a plethora of order parameters related to the SO density wave formations which break symmetry in different ways. Among them, we rule out those parameters which render gapless states by using the symmetry arguments (see {\it SI}): All four order parameters obey $\mathcal{I}$ symmetry, while only $M^y$ term is even under $\mathcal{TR}$, because it is the product of two odd terms ${\bm \tau}^y$ and $\bm{\sigma$} (we drop the superscript `$y$' henceforth). This is the only term which commences a finite gap opening if the translational or rotational symmetry is spontaneously broken. We have shown in {\it SI} that there exists a considerably large parameter space of the coupling constant `$g$' of the coupling constant where this order parameter dominates.

Eq.~\ref{eq:op} implies that spin and orbital orderings occur simultaneously along the `hot-spot' direction ${\bm Q}_h$, as illustrated in Fig.~\ref{fig2}{\bf a}. It propagates along ${\bm Q}^x_h$ or ${\bm Q}^y_h$ directions with alternating signs (particle-hole pairs) to commence a SO density wave. The resulting Hamiltonian breaks the four-fold rotational symmetry down to a two-fold one $\mathcal{C}_2$, and gives rise to a so-called spin-orbit `smectic' state which breaks both translational and $C_4$ symmetry.\cite{nematic1} The present $\mathcal{TR}$ invariant SO order parameter is inherently distinct from any spin or orbital or even interorbital spin-density wave order which break $\mathcal{TR}$ symmetry. This criterion also rules out any similarly between our present spin-orbit smectic state with the spin-nematic phase\cite{SN} or spin-liquid state\cite{Pepin}. Furthermore, the present order parameter is different from $\mathcal{TR}$ invariant `hybridization wave' (between $f$ and $d$ orbitals of same spin), or charge density wave or others,\cite{Santini,Kotliar} as SO order involves flipping of both orbital (between split $f$ orbitals that belong to $\Gamma_6$ symmetry) and spin simultaneously. Taking into account the band-structure information that $\bm{Q}_{h}$ represents the interband nesting, it is instructive to focus on only $\bm{b}_{12}(\bm{k})$ component (thus the subscript `12' is eliminated hereafter). Therefore, the SO density wave does not introduce a spin or orbital moment, but a polarization in the total angular momentum $\delta m_J$=$\pm2$ [for the ordering between $\frac{3}{2}$(-$\frac{3}{2}$) and -$\frac{1}{2} $($\frac{1}{2}$)].

The $\bm{b}$ vector belongs to the same irreducible point group representation, $E_g$, of the crystal with odd parity, and can be defined by $|\bm{b}({\bm k})|$=$2ig\Delta^x\sin{k_xa}$, or $2ig\Delta^y\sin{k_ya}$ for the wavevectors $\bm{Q}^x_{h}=(1\pm0.4,0,0)$, or $\bm{Q}^y_{h}=(0,1\pm 0.4,0)$, respectively. The mean-field Hamiltonian for the HO state within an effective two band model reduces to the general form $H_{MF}$=$H_0$+$H_{SODW}$, where the particle-hole coupling term is
\begin{equation}
H_{SODW} = 2i\sum_{{\bm k}}\sum_{\mu=x,y} \Delta^{\mu}\sin{(k_\mu a)}f^{\dag}_{\bm{k},1,\uparrow}f_{\bm{k}+\bm{Q}_h^{\mu},2,\downarrow} +h.c.
\label{eq:intHam}
\end{equation}
In the Nambu representation, it is obvious that the HO term merely adds a mass term to the $d_{12}^y$ term defined above. At the band-crossing points located where $|{\bm d}_{12}^2|$-$|{\bm d}_{11}^2|$=0, a gap opens by the value of $|{\bm b}({\bm k})|^2$. Figure~\ref{fig2} demonstrates the development of the quasi-particle structure in the HO state. The band progression and the associated gap opening is fully consistent with the angle-resolved photoemission spectroscopy (ARPES) observations\cite{ARPES,ARPES_FS}. The scanning tunneling microscopy and spectroscopic (STM/S)\cite{STM_Yazdani,STM_Davis} fingerprints of the gap opening in the density of state (DOS) is also described nicely within our calculations, see Fig.~\ref{fig2}{\bf c}.

\section{Discussion}
\vskip0.25cm
\noindent
The SO moment is $J^z_{\tau\tau^{\prime}\sigma\sigma^{\prime}}({\bm q},\mathcal{T})=\sum_{\bm k} f_{{\bm k},\tau,\sigma}^{\dag}(\mathcal{T})[{\bm \tau}_{\tau\tau^{\prime}}^{\mu}{\bm \sigma}_{\sigma\sigma^{\prime}}^{z}]f_{{\bm k}+{\bm q},\tau^{\prime},\sigma^{\prime}}(\mathcal{T})$, where $\mathcal{T}$ is imaginary time. Introducing simplified indices $\alpha,\beta=\tau\tau^{\prime}\sigma\sigma^{\prime}$, the correlation function of $J^z_{\alpha}$ vector can be defined as $\chi^{zz}_{\alpha\beta}({\bm q},\mathcal{T})=\frac{1}{N}\left\langle T_{\mathcal{T}}J^z_{\alpha}({\bm q},\mathcal{T})J^z_{\beta}(-{\bm q},0)\right\rangle$, where $T_{\mathcal{T}}$ is normal time-ordering. Our numerical calculation of the $\chi^{zz}_{\alpha\beta}({\bm Q}_h,\omega)$ within random-phase approximation (RPA) yields an inelastic neutron scattering (INS) mode with enhanced intensity at $\bm{Q}_{h}$ near $\omega_Q\sim 4.7$~meV below $T_{h}$ as shown in Fig.~1{\bf d}. INS data (symbols) at a slightly large momentum agrees well with our calculation, however, a polarized INS measurement will be of considerable value to distinguish our proposed $\mathcal{TR}$ invariant mode from any spin-flip and elastic background.\cite{polarizedINS} A $T$-dependent study of the INS mode also reveals that this mode becomes strongly enhanced at $Q_h$ rather than at the commensurate one below $T_h$.\cite{INS}

One way to characterize the nature of a phase transition is to determine the temperature evolution of the gap value. Our computed self-consistent values of the mean-field gap $\Delta(T)$ agree well with the extracted gap values from the STM spectra\cite{STM_Yazdani} [see {\it inset} to Fig.~2{\bf c}]. In general, the entropy loss at a mean-field transition is given by\cite{ChandraColeman} $\Delta S\sim k_B\frac{2\Delta}{\xi_F}$, where $\Delta$ is the HO gap and $\xi_F$ is the Fermi energy of the gapped state. At HO the Fermi energy $\xi_F\approx\sqrt{\frac{1}{2}\left[E_1({\bm k}_F)-E_2({\bm k}_F)\right]^2 + \Delta^2}$,  where the two linearly dispersive bands near the Fermi level yields $E_i({\bm k}_F)\approx \hbar{\bar v}_{iF}k_{iF}$. Using the measured Sommerfield coefficient $\gamma$=180~mJmol$^{-1}$K$^{-2}$, compared to its linear expansion of $\gamma_0$=50~mJmol$^{-1}$K$^{-2}$, we obtain the mass renormalization factor $Z^{-1}=\gamma/\gamma_0=3.6$. This gives ${\bar v}^i_F=Zv^i_F$. For the two bands that participate in the HO gap opening, we get $v_{1F}$=0.11$a/\hbar$ in eV at $k_{1F}$=0.5$\pi/a$ and $v_{2F}$=0.58$a/\hbar$ in eV at $k_{2F}$=0.3$\pi/a$ from Fig.~1{\bf a}. Using the experimental value of $\Delta$=5meV,\cite{STM_Yazdani,STM_Davis} we obtain $\Delta S\sim0.28k_B\ln{2}$, which is close the experimental value of $0.3k_B\ln{2}$.\cite{Cv_Palstra}

\begin{figure}[th]
\hspace{-0cm}
\rotatebox{0}{\scalebox{.6}{\includegraphics{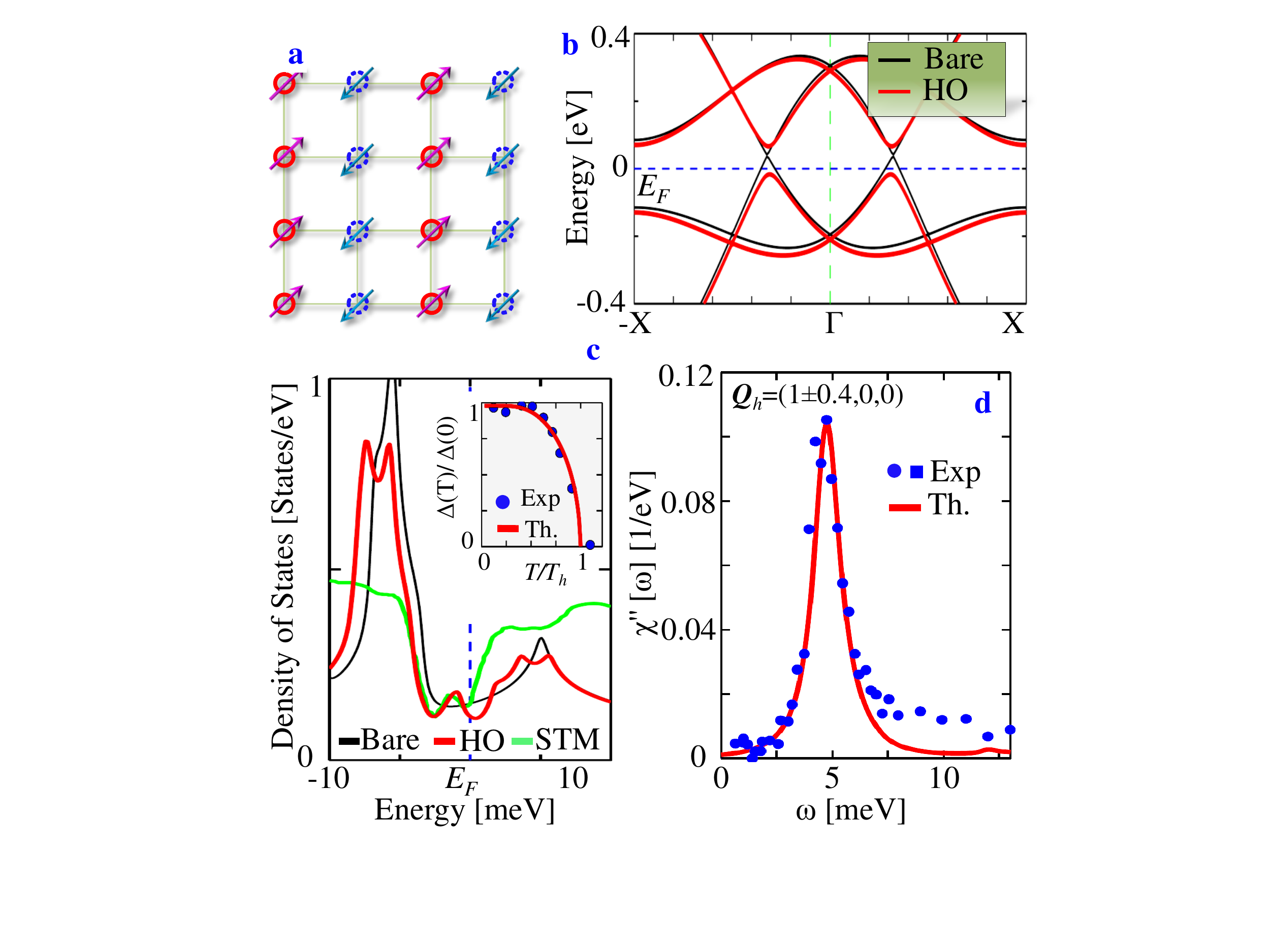}}}
\caption{{\bf Spin-orbit density wave and the hidden-order gap opening.} {\bf a}, A typical form of the staggered SO order is schematically described for an illustrative case of commensurate wavevector. The solid and dashed circles encode two opposite orbitals, $\tau=\pm$, where the associated arrows depict their `pseudospins' $\sigma$. Both $\bm{\tau}$ and $\bm{\sigma}$, representing orbital and spin respectively, individually break $\mathcal{TR}$ symmetry, while their product remains $\mathcal{TR}$ invariant. {\bf b}, Model dispersions of the $|\pm\frac{1}{2}\rangle$, and $|\pm\frac{3}{2}\rangle$ subbands plotted along the axial direction. Black and red lines give dispersion before and after including the HO gap, respectively. An artificially large value of $\Delta$=50~meV is chosen here to clearly explicate the momentum dependence of the modulated SO gap opening. {\bf c}, Modifications of DOS upon entering into the HO phase are compared with measured DOS in the STM experiment (green line).\cite{STM_Davis} Note that the experimental data is subtracted from the background spectrum at $T$$>$$T_{h}$, which helps highlight the appearance of multiple structures in the DOS spectrum at the HO state. Here the gap magnitude $\Delta(0)$=5~meV, obtained at a coupling strength of $g=27$~meV, see SI.  {\it Inset:} The self-consistent value of $\Delta(T)$ exhibits the mean-field behavior of the HO gap, in consistent with experiments.\cite{STM_Yazdani} We obtain $T_{h}=22$~K which is larger than the experimental value of $T_{h}=17.5$~K. However, recently it has been pointed out that there exists a `pseudogap' above the HO state,\cite{pseudogap} which presumably reduces the mean-field temperature scale. {\bf d}, RPA result of SO correlation function at $g=$28.4~meV shows a resonance peak at $\omega_Q=4.7$~meV at $Q_h$, in good agreement with experimental data.\cite{INS_JPSC,Broholm}}
\label{fig2}
\end{figure}

We now evaluate the topological invariant index of interacting Hamiltonian in Eq.~\ref{eq:intHam} to demonstrate that HO gap opening in URu$_2$Si$_2$ also induces topological phase transition. To characterize the topological phenomena, we recall the Fu-Kane classification scheme\cite{FuKane} which implies that if a time-reversal invariant system possess an odd value of $Z_2$ invariant index, the system is guaranteed to be topologically non-trivial. $Z_2$ index is evaluated by the time-reversal invariant index $\nu_i=\pm 1$, if defined, for all filled bands as $Z_2=\nu_1\nu_2 ... \nu_n$, where $n$ is the total number of orbitals in the Fermi sea. A more efficient method of determining the topological phase is called the adiabatic transformation scheme used earlier in realizing a large class of topological systems, especially when $Z_2$ calculation is difficult.\cite{DasTI} In this method, the non-trivial topological phase of a system can be realized by comparing its band-progression with respect to an equivalent trivial topological system. URu$_2$Si$_2$ is topologically trivial above the HO state, i.e. $Z_2^0=+1$. The gap opening makes the top of the valence band (odd parity) to drop below $E_F$ as shown in Fig.~\ref{fig2}{\bf b}. Thereby, an odd parity gained in the occupied level endows the system to a non-trivial topological metal. To see that we evaluate the topological index for the HO term as $\nu_{ho} = \int d{\bm k} \Omega({\bm k})$, where the corresponding Berry curvature can be written in terms of ${\bm b}$-vector as $\Omega=\hat{b}\cdot\left(\frac{\partial \hat{b}}{\partial k_x}\times\frac{\partial \hat{b}}{\partial k_y}\right)$ for each spin with $\hat{b}={\bm b}/|{\bm b}|$. Due to the odd parity symmetry of ${\bm b}$, it is easy to show that $\nu_{ho}=-1$ which makes the total $Z_2$ value of the HO phase to be $Z_2=Z_2^0\times\nu_{ho}=-1$, and hence we show that hidden order gapping is a topologically non-trivial phase. The consequence of a topological bulk gap is the presence of surface states.\cite{FuKane,DasTI} In our present model, we expect two surface states of opposite spin connecting different orbitals inside the HO gap. As the system is a weak-topological system, the surface states are unlikely to be topologically protected. The spin-orbit locking of these states can be probed by ARPES using circular polarized incident photon which will be a definite test of this postulate.

The HO gap is protected from any $\mathcal{TR}$ invariant perturbation such as pressure (with sufficient pressure the HO transforms into the LMAF phase), while $\mathcal{TR}$ breaking perturbation such as magnetic field will destroy the order. Remarkably, these are the hallmark features of the HO states,\cite{SN,Matsuda,HarrisonQCP} which find a natural explanation within our SO density wave order scenario. In what follows, the magnetic field will destroy the HO state even at $T=0$~K, that means at a quantum critical point (QCP) as the HO is a spontaneously broken symmetry phase.\cite{sachdev} However, due to the finite gap opening at the HO state, it requires finite field to destroy the order. The thermodynamical critical field can be obtained from\cite{pathria} $\Delta=\langle\chi_Q(\omega_{res})\rangle B_c^2$, where $B_c$ is the critical field and $\chi_Q(\omega_{res})\approx2\Delta\alpha^2\rm{tanh}(\Delta/2k_BT_h)/\omega^2_{res}$ at the resonance mode that develops in the HO state. $\alpha=g\mu_B|\langle \delta m_J\rangle|=2g\mu_B$ and the bare `g'-factor $g=0.8$. Substituting $\omega_{res}=4.7$meV, we get the location of the QCP at $B\approx38$~T, which is close the experimental value of $B=34$~T.\cite{HarrisonQCP}

Broken symmetry FS reconstruction leads to enhanced Nernst signal.\cite{Nernst} For the case of broken symmetry spin-orbit order, we expect to generate spin-resolved Nernst effect which can be measured in future experiments to verify our proposal.\cite{spinNernst}

In summary, we proposed a novel SO density wave order parameter for the HO state in URu$_2$Si$_2$. Such order parameter is $\mathcal{TR}$ symmetry invariant. We find no fundamental reason why such order parameter cannot develop in other systems in which both electronic correlation and SO of any kind are strong. Some of the possible materials include heavy fermion systems, Iridates,\cite{iridates} SrTiO$_3$ surface states,\cite{santander} SrTiO$_3$/LiAlO$_3$ interface,\cite{oxide} Half-Heusler topological insulator\cite{DasTI}, and other $d$- and $f$-electron systems with strong SO. In particular, a Rashba-type SO appears due to relativistic effect in two-dimensional electron system yielding helical FSs. In such systems, the FS instability may render similar SO density wave, and the resulting quasiparticle gap opening is observed on the surface state of BiAg$_2$ alloys even when the spin-degeneracy remains intact.\cite{BiAg2} Furthermore, recent experimental findings of quasiparticle gapping in the surface state of topological insulator due to quantum phase transition even in the absence of time-reversal symmetry breaking can also be interpreted as the development of some sort of spin orbit order.\cite{TIgap}

{\bf Acknowledgments}\\
The author thanks A. V. Balatsky, M. J. Graf, A. Bansil, R. S. Markiewicz, T. Durakiewicz, J.-X. Zhu, P. M. Oppeneer, J. Mydosh, P. W\"olfle, and J. Haraldsen for useful discussions. Work at the Los Alamos National Laboratory was supported by the U.S. DOE under contract no. DE-AC52-06NA25396 through the Office of Basic Energy Sciences and the UC Lab Research Program, and benefited from the allocation of supercomputer time at NERSC.

\clearpage

\clearpage

\clearpage
\section{Supplementary Information (SI) for ``Spin-orbit density wave induced hidden topological order in URu$_2$Si$_2$"}

In the main text, we have provided an effective two band model which is relevant for the study of hidden-order (HO) gap structure coming from Fermi surface (FS) instability. Here, we expand on how such an effective Hamiltonian is deduced. The spin-orbit splitting state is, in general, studied within either ${\bm L}.{\bm S}$ coupling or $j-j$ coupling approaches. In the former case, the total $\bm{L}$ and $\bm{S}$ are formed due to strong Hund's coupling prior to the formation of spin-orbit coupled eigenstates. Such process mainly occurs in insulating compounds with localized $f$ states.\cite{DzeroPRL} However, in actinides the spin-orbit (SO) coupling is stronger than the Hund's coupling.\cite{actinidesSO} Therefore, total angular momentum ${\bm J}={\bm L} + {\bm S}$ is the good quantum number for this state. In such systems, a SO density wave order in the two-particle channel can arise at some critical value of the coupling constant $g$, by taking advantage of any instability, such as FS `nesting' (shown in the main text), even when the time-reversal symmetry remains invariant. The coupling $g$ can be related to some form of `screened' interorbital Coulomb term. Physically, SO density wave is different from a spin density wave because here spin-flip occurs between two different orbitals without breaking time-reversal symmetry. Therefore, we desire to study a Hamiltonian:
\begin{eqnarray}
H = H_0 +  H_{SODW},
\end{eqnarray}
where $H_0$ is the non-interacting part and $H_{SODW}$ is the SO density wave term. In the `pseudospin' basis introduced in the main text $\hat{\Psi}^{\dag}({\bm k})=(f^{\dag}_{{\bm k},\frac{1}{2},\sigma},~f^{\dag}_{{\bm k},\frac{3}{2},\sigma},f^{\dag}_{{\bm k},\frac{1}{2},\bar{\sigma}},~f^{\dag}_{{\bm k},\frac{3}{2},\bar{\sigma}})$ ($\bar{\sigma}=-\sigma$), the representation of the symmetry operations for URu$_2$Si$_2$ system which belong to the $D_{4h}$ symmetry is: time-reversal symmetry $\mathcal{TR}=\mathcal{K}\cdot i{\bm \sigma}^y\otimes{\bm \tau}^0$, inversion symmetry $\mathcal{I}={\bm \tau}^z\otimes\bm{\tau}^0$, four-fold rotation symmetry around the $z$ axis $\mathcal{C}_4=\exp{[i(\pi/4){\bm \sigma}^z\otimes\bm{\tau}^0]}$ and the two reflection symmetries $\mathcal{P}_{x/y}=i{\bm \sigma}^{x/y}$ which map $x\rightarrow -x$ (where $x$ is in $\Gamma$-X direction) and $y\rightarrow -y$ (where $y$ is orthogonal to $\Gamma$-X direction), respectively. Here, $\mathcal{K}$ is complex conjugation operator, and $\bm{\sigma}^{x,y,z}$ and $\bm{\tau}^{x,y,z}$ depict the two-dimensional Pauli matrices in the `pseudospin' and orbital space, respectively where $\tau^0$ is the  unitary matrix.

Each symmetry operation transforms the time-reversal invariant $f$-electron field as:
\begin{eqnarray}
\mathcal{I}\hat{\Psi}(k_x,k_y)&\rightarrow& -\bm{\tau}^0\hat{\Psi}(-k_x,-k_y),\label{symm1}\\
\mathcal{C}_4\hat{\Psi}(k_x,k_y)&\rightarrow& i\bm{\tau}^y\hat{\Psi}(-k_y,k_x),\label{symm2}\\
\mathcal{P}_{x/y}\hat{\Psi}(k_x,k_y)&\rightarrow& \mp\bm{\tau}^{z}\hat{\Psi}(\mp k_x,\pm k_y).\label{symm3}
\end{eqnarray}
These symmetry operations imply that the spin-orbit coupled actinide $f$-state is an odd-parity wavefunction. Using these symmetry properties, the non-interacting Hamiltonian is deduced in the main text, using standard procedure, see for example Refs.~\onlinecite{SCZSHE,SCZTFT}.

\section{Hidden order parameter}

Considering the non-interacting FS nesting at ${\bm Q}_h=(1\pm0.4,0,0)$ for the hidden order state as demonstrated in the main text, we expand the above-mentioned `pseudospinor' in the Nambu notation as $\hat{\Psi}({\bm k})=\left(\hat{\Psi}({\bm k}), \hat{\Psi}({\bm k}+{\bm Q}_h) \right)$. In this notation, the modulated SOC interaction term can be written in general as
\begin{eqnarray}
H_{SODW} = \sum_{\mu\nu} g^{\mu\nu} : \left[\hat{\Psi}^{\dag}(\bm{k})\Gamma^{\mu\nu}\hat{\Psi}(\bm{k}+\bm{Q}_h)\right]^2:,
\end{eqnarray}
where $\mu,\nu\in\{0,x,y,z\}$. The symbol $::$ represents normal ordering. Here $g$ is the coupling constant discussed latter and $\Gamma^{\mu\nu}={\bm \tau}^{\mu}\otimes\bm{\sigma}^{\nu}$, $\bm{\tau}$ and $\bm{\sigma}$  represent Pauli matrices in orbital and spin basis, respectively. Absorbing $g$ and $\Gamma$ into one term we define the mean-field order parameter
\begin{eqnarray}
M^{\mu\nu} &=& g^{\mu\nu}(\bm{k})\left\langle\hat{\Psi}({\bm k})^{\dag}[{\bm \tau}^{\mu}\otimes{\bm \sigma}^{\nu}]\hat{\Psi}({\bm k}+{\bm Q}_h)\right\rangle.\\
& =& g^{\mu\nu}(\bm{k})\left\langle f_{{\bm k},\tau,\sigma}^{\dag}[{\bm \tau}_{\tau\tau^{\prime}}^{\mu}\otimes{\bm \sigma}_{\sigma\sigma^{\prime}}^{\nu}]f_{{\bm k}+{\bm Q}_h,\tau^{\prime},\sigma^{\prime}}\right\rangle.
\label{Mvector0}
\end{eqnarray}
Here $\tau,\tau^{\prime}$ and $\sigma,\sigma^{\prime}$ (not in bold font) are the components of the ${\bm \tau}^{\mu}$ and ${\bm \sigma}^{\nu}$ matrices, respectively. Without any loss of generality we fix the spin orientation along $z$-directions ($\nu=z$).  Therefore, we drop the index $\nu$ henceforth. Furthermore we define the gap vector as
\begin{eqnarray}
{\bm b}^{\mu}_{\tau\tau^{\prime}}({\bm k})=\Delta_{\tau\tau^{\prime}}^{\mu}({\bm k}){\bm \tau}_{\tau\tau^{\prime}}^{\mu}, 
\label{bvector}
\end{eqnarray}
where momentum dependence of the gap function transforms according to the same irreducible representation of the point-group symmetry ($g$ is absorbed in the gap function defined below). With these substitutions, we obtain the final result for the order parameter as
\begin{eqnarray}
M^{\mu} &=& \left\langle \sum _{\tau\tau^{\prime}\sigma\sigma^{\prime}} f_{{\bm k},\tau,\sigma}^{\dag}\left[{\bm b}_{\tau\tau^{\prime}}^{\mu}({\bm k}){\bm \sigma}_{\sigma\bar{\sigma}}^z\right]f_{{\bm k}+{\bm Q}_h,\tau^{\prime},\bar{\sigma}}\right\rangle.
\label{Mvector}
\end{eqnarray}
 For the unidirectional modulation vector $\bm{Q}_h$, all components of $M^{\mu}$ break $C_4$ rotational symmetry. All the symmetry properties of the order parameters are given in Table I. $M^0,M^x$ and $M^z$ break time-reversal and thus are ruled out as the hidden-order state is arguably does not exhibit any time-reversal symmetry breaking.\cite{Kotliar,Sasha,Santini}. Although some evidences for the time-reversal symmetry breaking are also present, but it is not well established if the measurements are done in single crystal where the time-reversal symmetry breaking LMAF state and time-reversal symmetry invariant HO state are not mixed. In microscopic sense, LMAF and HO state are separated by phase transition, and thus both phases cannot inherit same broken symmetry.

\begin{table}[h]
\centering
\begin{tabular}{|c|c|c|c|c|}
\hline \hline
& $M^0$ & $M^x$ & $M^y$ &$M^z$ \\
\hline
$\mathcal{TR}$ & - & - & + & -\\
$\mathcal{C}_4$ & - & - & - & -\\
$\mathcal{I}$ & + & + & + & + \\
$\mathcal{P}_{x/y}$ & + & - & - & + \\
\hline \hline
\end{tabular}
\caption{Symmetry properties of various density wave parameters. `+' (`-') represent `even'
(`odd') parity of the order parameter under the corresponding symmetry operations.}
\label{Tab:symm}
\end{table}

On the other hand, $M_2$ is even under time-reversal but odd under parity, and thus adds a mass term to the Hamiltonian which opens a gap at the nested portion of the Fermi surface. This is the term that represents modulated spin-orbit coupling. A trivial check can be performed by explicitly writing down the the spin-orbit coupling term $\sum_i(-1)^{i_x}{L}_{i,z}{S}_{i,z}\propto f_{i,+,\sigma}^{\dag}f_{i,+,\sigma^{\prime}}-f_{i,-,\sigma}^{\dag}f_{i,-,\sigma^{\prime}}=2if_{i,\frac{1}{2},\sigma}^{\dag}f_{i,\frac{3}{2},\sigma^{\prime}}+h.c.$, where we substituted $f_{i,\pm,\sigma}=(f_{i,\frac{1}{2},\sigma}\pm if_{i,\frac{3}{2},\sigma})$.

\begin{figure}[top]
\hspace{-0cm}
\rotatebox{0}{\scalebox{.5}{\includegraphics{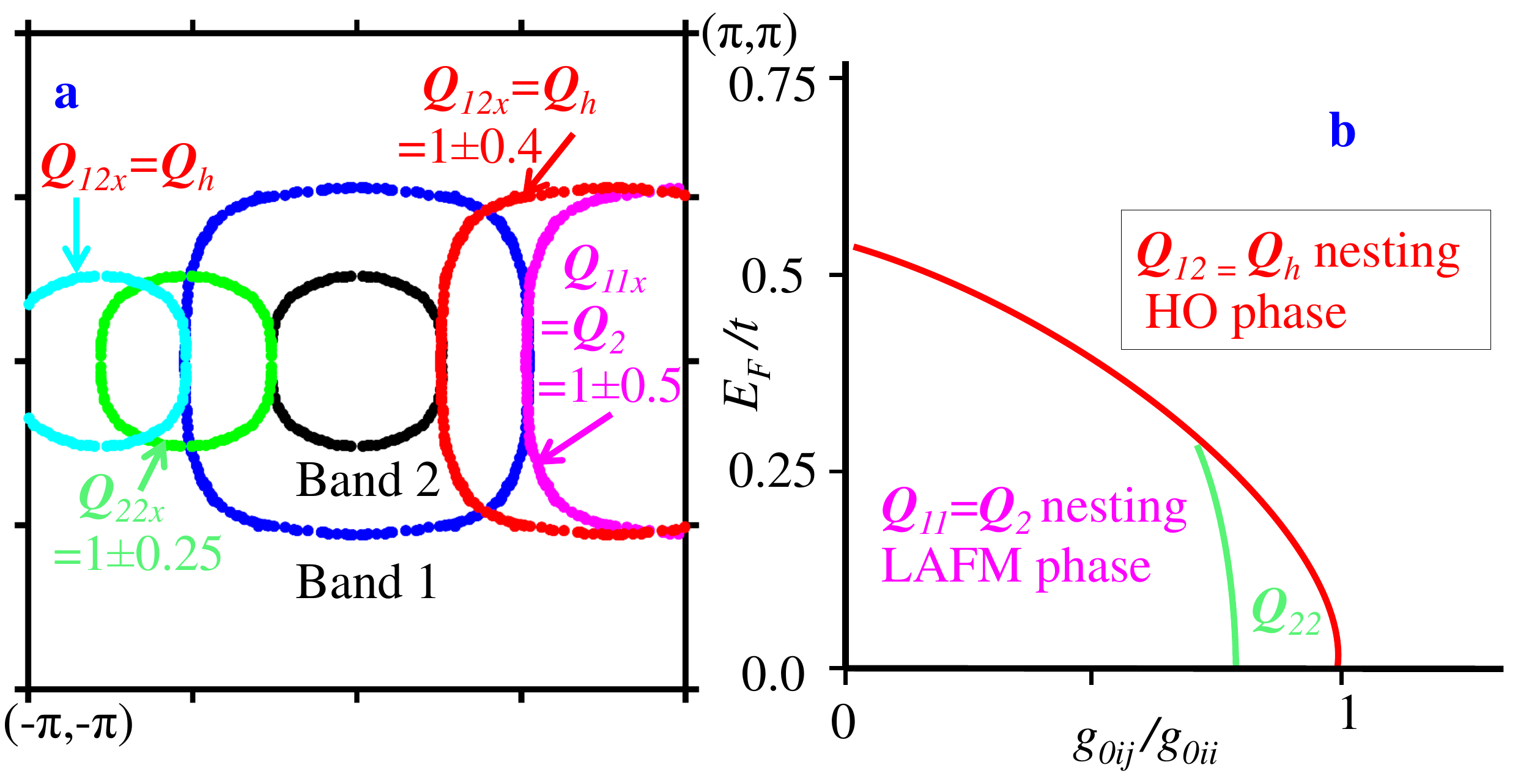}}}
\caption{{\bf a}, Black and blue symbols are {\it ab-initio} FSs, as shown in Fig.~1 of main text, but plotted here only two relevant bands. Two large square FSs for band 1 are shifted along the $x-$direction by nesting vectors ${\bm Q}_{11}={\bm Q}_2$ and ${\bm Q}_{12}$, while the two small square FSs for band 2 are shifted along $-x$ direction. Good nestings for all three vectors are observed, but the intraband one are protected by strong SOC, where ${\bm Q}_{12}={\bm Q}_h$ is the hidden order nesting that opens a gap. {\bf b}, Phase diagram as a function of intra-orbital ($g_{0ij}=g_{012}=g_{021}$) and inter-orbital coupling constant ($g_{0ii}=g_{011}=g_{022}$) and nearest neighbor hopping ($t$) with respect to the Fermi energy ($E_F$) for all thee nesting instabilities. We find that for smaller inter-orbital coupling than the intra-orbital one, the hidden order parameter arises if $E_F/t$ ration is large. With large $t$ (proportional to pressure), the LMAF phase at the commensurate nesting wins. This result is consistent, at least, in qualitative level, with the phase diagram of URu$_2$Si$_2$.\cite{phase_diagram} }
\label{sfig1}
\end{figure}
With this definition, the total Hamiltonian in the reduced Brillouin zone can be written as
\begin{eqnarray}
H &=& \sum_{{\bm k},\tau,\tau^{\prime},\sigma}^{\prime} \left( h_{\tau\tau^{\prime}} ({\bm k}) f_{{\bm k},\tau,\sigma}^{\dag}f_{{\bm k},\tau^{\prime},\bar{\sigma}} \right.  \nonumber\\
&&\left. ~~~~~~~~~+ h_{\tau\tau^{\prime}} ({\bm k}+{\bm Q}_h) f_{{\bm k}+{\bm Q}_h,\tau,\sigma}^{\dag}f_{{\bm k}+{\bm Q}_h,\tau^{\prime},\bar{\sigma}} \right) \nonumber\\
&&-\sum_{{\bm k},\mu,\tau,\tau^{\prime},\sigma}^{\prime} \left(M^{\mu}({\bm k})f_{{\bm k},\tau,\sigma}^{\dag}[{\bm \tau}_{\tau\tau^{\prime}}^{\mu}\otimes{\bm \sigma}_{\sigma\sigma^{\prime}}^{\nu}]f_{{\bm k}+{\bm Q}_h,\tau^{\prime},\sigma^{\prime}} \right.\nonumber\\
&&\left. ~~~~~~~~~ + M^{\mu}({\bm k}+{\bm Q}_h)f_{{\bm k}+{\bm Q}_h,\tau,\sigma}^{\dag}[{\bm \tau}_{\tau\tau^{\prime}}^{\mu}\otimes{\bm \sigma}_{\sigma\sigma^{\prime}}^{\nu}]f_{{\bm k},\tau^{\prime},\sigma^{\prime}}\right) \nonumber\\
&&+ \sum_{{\bm k},\mu} M^{\mu}({\bm k})M^{\mu}({\bm k}+{\bm Q}_h)/g^{\mu}.
\end{eqnarray}
Where $h$ is the non-interacting part of the Hamiltonian defined in Eq.~2 of the main text. The prime over a summation implies that the summation is performed in the reduced Brillouin zone. Diagonalizing the above Hamiltonian, we deduce the quasiparticle states are: $E^{\tau\sigma}_{\bm k,\nu}=\xi^{\tau\sigma}_{{\bm k}+}+\nu E^{\tau\sigma}_{\bm k,0}$, and $E^{\tau\sigma}_{\bm k,0}=
\sqrt{(\xi^{\tau\sigma}_{{\bm k}-})^2+|\Delta_{\tau\tau^{\prime}}|^2}$, where $\xi^{\tau\sigma}_{{\bm k}\pm}=(E^{\tau\sigma}_{\bm k}\pm E^{\tau\sigma}_{{\bm k}+{\bm Q}})/2$, and
\begin{eqnarray}
E^{\tau\sigma}(\bm{k})=\epsilon({\bm k})+\tau\sqrt{\sum_{\mu}|{\bm d}^{\mu}_{12}({\bf k})|^2}+\sigma\sqrt{\sum_{\mu}|{\bm d}^{\mu}_{11}({\bm k})|^2}\nonumber\\
\end{eqnarray}
from Eq.~3 of the main text. $\nu,\nu^{\prime}=\pm$. The corresponding coherence factors are
\begin{eqnarray}
u^{\tau\sigma}_{\bm k} (v^{\tau\sigma}_{\bm k})=\frac{1}{2}\left(1\pm \frac{\xi^{\tau\sigma}_{{\bm k}-}}{E^{\tau\sigma}_{\bm k,0}}\right).
\end{eqnarray}

\section{Self-consistent gap equation}
We can easily derive the self-consistent gap equation from Eq.~\ref{Mvector} using Bogolyubov treatment. We substitute the fermion operator $f$ in terms of a Bogolyubov operators as
\begin{eqnarray}
f_{\bm{k},\tau\sigma}&=&u^{\tau\sigma}_{\bm k}\gamma_{\bm{k},\tau\sigma}-v^{\tau\sigma}_{\bm k}\gamma_{\bm{k}+{\bm Q}_h,\tau\sigma}\nonumber\\
f_{\bm{k}+{\bm Q}_h,\tau\sigma}&=&v^{\tau\sigma}_{\bm k}\gamma_{\bm{k},\tau\sigma}+u^{\tau\sigma}_{\bm k}\gamma_{\bm{k}+{\bm Q}_h,\tau\sigma}.
\end{eqnarray}
From Eqs.~\ref{Mvector0},~\ref{bvector}, and \ref{Mvector}, we get the gap function as
\begin{eqnarray}
\Delta^{\mu}_{\tau\tau^{\prime}}&=&g^{\mu}_{\tau\tau^{\prime}}\sum_{{\bm k},\sigma} \left\langle f_{{\bm k}+{\bm Q}_h,\tau\sigma}^{\dag}f_{{\bm k},\tau^{\prime}\bar{\sigma}}\right\rangle,\\
&=&g^{\mu}_{\tau\tau^{\prime}}\sum_{{\bm k},\sigma}u^{\tau\sigma}_{\bm k}v^{\tau^{\prime}\bar{\sigma}}_{\bm k}\left[ \left\langle \gamma_{\bm{k},\tau\sigma}^{\dag}\gamma_{\bm{k},\tau^{\prime}\bar{\sigma}} \right\rangle \right.\nonumber\\
&&~~~~~~~~~\left.- \left\langle \gamma_{\bm{k}+{\bm Q}_h,\tau}^{\dag}\gamma_{\bm{k}+{\bm Q}_h,\tau^{\prime}\bar{\sigma}} \right\rangle \right],\nonumber\\
&=&g^{\mu}_{\tau\tau^{\prime}}\sum_{{\bm k},\sigma}u^{\tau\sigma}_{\bm k}v^{\tau^{\prime}\bar{\sigma}}_{\bm k}\left[ n_{f,\tau\sigma}^+({\bm k})-n_{f,\tau^{\prime}\bar{\sigma}}^-({\bm k})\right],\nonumber\\
&=&\frac{1}{2}g^{\mu}_{\tau\tau^{\prime}}\sum_{{\bm k},\sigma}\frac{\Delta^{\mu}_{\tau\tau^{\prime}}({\bm k})}{E_{\bm{k}}^{\tau\sigma}}\left[n_{f,\tau\sigma}^+({\bm k})-n_{f,\tau^{\prime}\bar{\sigma}}^-({\bm k})\right].\nonumber\\
\end{eqnarray}
Here we have substituted $\langle\gamma_{\bm{k},\tau\sigma}^{\dag}\gamma_{\bm{k},\tau\sigma}\rangle=n_{f,\tau\sigma}^{\nu}({\bm k})$, where $n$ is the Fermi function.
%
%
%
%
We find that $g_{12}=27$~meV gives $\Delta=$5~meV in consistent with experiments. For this temperature independent value of $g_{12}$, we obtain $T_h=22$K, which is higher than the experimental value of $T_h=17.5$~meV. Possible reason for overestimating the value of $T_h$ are the neglect of quantum fluctuation, disorder which can reduce its value. In this context, it can be noted that recently, a `pseudogap' phase upto 20~K is marked from experimental features,\cite{pseudogap} which arguably suggests that there are indeed fluctuations present above $T_h$.

\section{spin-orbit correlation function}

The non-interacting single-particle Green's function for the Hamiltonian given in Eq.~1 above is defined as $G_{\tau\tau^{\prime}\sigma\sigma^{\prime}}({\bm k},i\omega_n)=-{1}/{\beta}\sum_n \langle T_{\mathcal{T}}f_{\bm{k}\tau\sigma}(\mathcal{T})f_{\bm{k}\tau^{\prime}\sigma^{\prime}}^{\dag}(0)\rangle e^{i\omega_n\mathcal{T}}$.
Here $\mathcal{T}$ is the imaginary time, $f_{\bm{k}\tau\sigma}(\mathcal{T})$ is the imaginary time evolution of the fermionic operator $f_{\bm{k}\tau\sigma}$, $n$ is the fermionic Matsubara frequency, and $\beta=1/k_BT$, where $k_B$ is Boltzmann constant. $T_{\mathcal{T}}$ gives normal time-ordering. The anomalous part of the Green's function is
%
%
$F_{\tau\tau^{\prime}\sigma\sigma^{\prime}}({\bm k},i\omega_n) = -\frac{1}{\beta}\sum_n \left\langle T_{\mathcal{T}}f_{\bm{k}\tau\sigma}(\mathcal{T})f_{(\bm{k}+{\bm Q})\tau^{\prime}\sigma^{\prime}}^{\dag}(0)\right\rangle e^{i\omega_n\mathcal{T}}.$
%
%
By Fourier transforming in the Matsubara frequency space, we obtain the explicit forms of two Green's functions as
\begin{eqnarray}
G_{\tau\tau^{\prime}\sigma\sigma^{\prime}}({\bm k},i\omega_n)&=& \sum_{\nu}R^{\nu}_{\tau\tau^{\prime}\sigma\sigma^{\prime}}({\bm k})\nonumber\\
&&\times\left(\frac{(u^{\tau\sigma}_{\bm k})^2}{i\omega_n - E^{\tau\sigma}_{{\bm k},\nu}} + \frac{(v^{\tau^{\prime}\sigma^{\prime}}_{\bm k})^2}{i\omega_n - E^{\tau^{\prime}\sigma^{\prime}}_{{\bm k},\nu}}\right)\\
F_{\tau\tau^{\prime}\sigma\sigma^{\prime}}({\bm k},i\omega_n)&=& \sum_{\nu}R^{\nu}_{\tau\tau^{\prime}\sigma\sigma^{\prime}}({\bm k}) u^{\tau\sigma}_{\bm k}v^{\tau^{\prime}\sigma^{\prime}}_{\bm k}\nonumber\\
&&\times \left(\frac{1}{i\omega_n - E^{\tau\sigma}_{{\bm k},\nu}} - \frac{1}{i\omega_n - E^{\tau^{\prime}\sigma^{\prime}}_{{\bm k},\nu}}\right).
\end{eqnarray}
The orbital overlap matrix-element $R_{\tau\tau^{\prime}\sigma\sigma^{\prime}}^{\nu}({\bm k})=\psi_{\tau\sigma}^{\nu}({\bm k})\psi_{\tau^{\prime}\sigma^{\prime}}^{\nu\dag}({\bm k})$, where $\psi$ is the eigenvector of the noninteracting Hamiltonian, projects the Green's function from band basis to the orbital one. However, to simplify our calculation, we assume that each block state $E^{\tau\sigma}$ corresponds to each orbital which makes $R=1$. By taking analytical continuation of the Matsubara frequency to the real frequency in the above Green's function, it is easy to show that the gap function $\Delta$ can be obtained by averaging the anomalous Green's function.

The general form of the polarization vector is
\begin{equation}
J^z_{\tau\tau^{\prime}\sigma\sigma^{\prime}}({\bm q},\mathcal{T})=\sum_{\bm k} f_{{\bm k},\tau,\sigma}^{\dag}(\mathcal{T}){\bm \tau}_{\tau\tau^{\prime}}^{\mu}{\bm \sigma}_{\sigma\sigma^{\prime}}^{z}f_{{\bm k}+{\bm q},\tau^{\prime},\sigma^{\prime}}(\mathcal{T}).
\label{Jz}
\end{equation}
 To simplify the notations, we define composite indices $\alpha,\beta=\tau\tau^{\prime}\sigma\sigma^{\prime}$, in which the correlation function of $J^z_{\alpha}$ vector can now be defined as
\begin{eqnarray}
\chi^{zz}_{\alpha\beta}({\bm q},\mathcal{T})=\frac{1}{N}\left\langle T_{\mathcal{T}}J^z_{\alpha}({\bm q},\mathcal{T})J^z_{\beta}(-{\bm q},0)\right\rangle.
\end{eqnarray}
Substituting $J^z_{\alpha}$ and then applying standard Wick's decomposition to the electron bracket\cite{Mahan} yields,
%
\begin{eqnarray}
\chi^{zz}_{\alpha\beta}({\bm q},ip_m)&=&-\frac{1}{N}\sum_{\bm{k},n}\left[G_{\alpha}({\bm k},i\omega_n)G_{\beta}({\bm k}+{\bm q},i\omega_n+ip_m)\right.\nonumber\\
&&~~~\left.-F_{\alpha}({\bm k},i\omega_n)F_{\beta}({\bm k}+{\bm q},i\omega_n+ip_m)\right\rangle].\\
&=&\frac{1}{N}\sum_{\bm{k}}S_{\alpha\beta}({\bm k},{\bm q})\\
&&~~~\times\left[\left(u_{\bm k}u_{{\bm k}+{\bm q}} + v_{\bm k}v_{{\bm k}+{\bm q}}\right)^2(\chi^{++}+\chi^{--})\right. \nonumber\\
&&~~~\left.+\left(u_{\bm k}v_{{\bm k}+{\bm q}} - v_{\bm k}u_{{\bm k}_{\bm q}}\right)^2(\chi^{+-}+\chi^{-+})\right].
\label{chi1}
\end{eqnarray}
Here $S_{\alpha\beta}({\bm k},{\bm q})=R_{\alpha}({\bm k})R_{\beta}({\bm k}+{\bm q})$ is the matrix-element term which projects the susceptibility from the band representation to the orbital one. $\chi^{\nu\nu^{\prime}}$ are the fermionic oscillator terms in the band space $\nu,\nu^{\prime}=\pm$:
\begin{eqnarray}
\chi^{\nu\nu^{\prime}}=-\frac{n^{\nu}_{f}({\bm k})-n^{\nu^{\prime}}_{f}({\bm k}+{\bm q})}{ip_m-E^{\nu}_{\bm k}-E^{\nu^{\prime}}_{{\bm k}+{\bm q}}}.
\label{chi0}
\end{eqnarray}
The momentum, energy, $\alpha$ indices on both sides of the above equation are implicit.

\subsection{Hidden order instability}
The divergence in the real-part of the $\chi_{\alpha\beta}({\bm q},\omega=0)$ indicates an instability due to the FS nestings between two orbitals $\alpha$ and $\beta$ (in the explicit notation $\chi_{11}$ means $\chi_{1111}$, etc). The two intraband nestings at $Q_{11}$ and $Q_{22}$ can give rise to spin-density wave or antiferromagnetism, if the time-reversal symmetry is broken. We believe that $Q_{11}$ is responsible for the LMAF phase. Our present interest is the interband one at $Q_{12}$, shown in supplementary Fig.~\ref{sfig1}({\it A}) which happens at $(1\pm0.4,0,0)$. Since in this case, both orbital and spin flip together, the time-reversal symmetry remains intact.

The corresponding critical value of the coupling constants $g_{\alpha\beta}$ at which a gap opening or an order parameter develops can be evaluated within random-phase approximation (RPA). The stoner criterion for an instable state implies that $1-\chi^{\prime}_{\alpha\beta}({\bm Q}_{\alpha\beta},0)g_{\alpha\beta}\ge0$, or $g_{\alpha\beta}\le1/\chi^{\prime}_{\alpha\beta}({\bm Q}_{\alpha\beta},0)$. Looking at the FS areas for each band,  shown in supplementary Fig.~\ref{sfig1}{\bf a}, we immediately see that $1/\chi^{\prime}_{11}({\bm Q}_{11},0)< 1/\chi^{\prime}_{22}({\bm Q}_{22},0)< 1/\chi^{\prime}_{12}({\bm Q}_{12},0)$, leading to a phase diagram shown in supplementary Fig.~\ref{sfig1}({\it B}). The present calculation does not incorporate the possible coexistence state between different phases. The phase diagram implies that there is a considerably large parameter space, where the ${\bm Q}_{12}={\bm Q}_h$ nesting dominates. We have not considered all possible phases except $M^y$ for $Q_{12}$ nestings discussed in Eq.~\ref{Mvector} above, because constrained by the symmetry arguments given in Table~I, others render gapless state.

\begin{figure}[top]
\hspace{-0cm}
\rotatebox{0}{\scalebox{.45}{\includegraphics{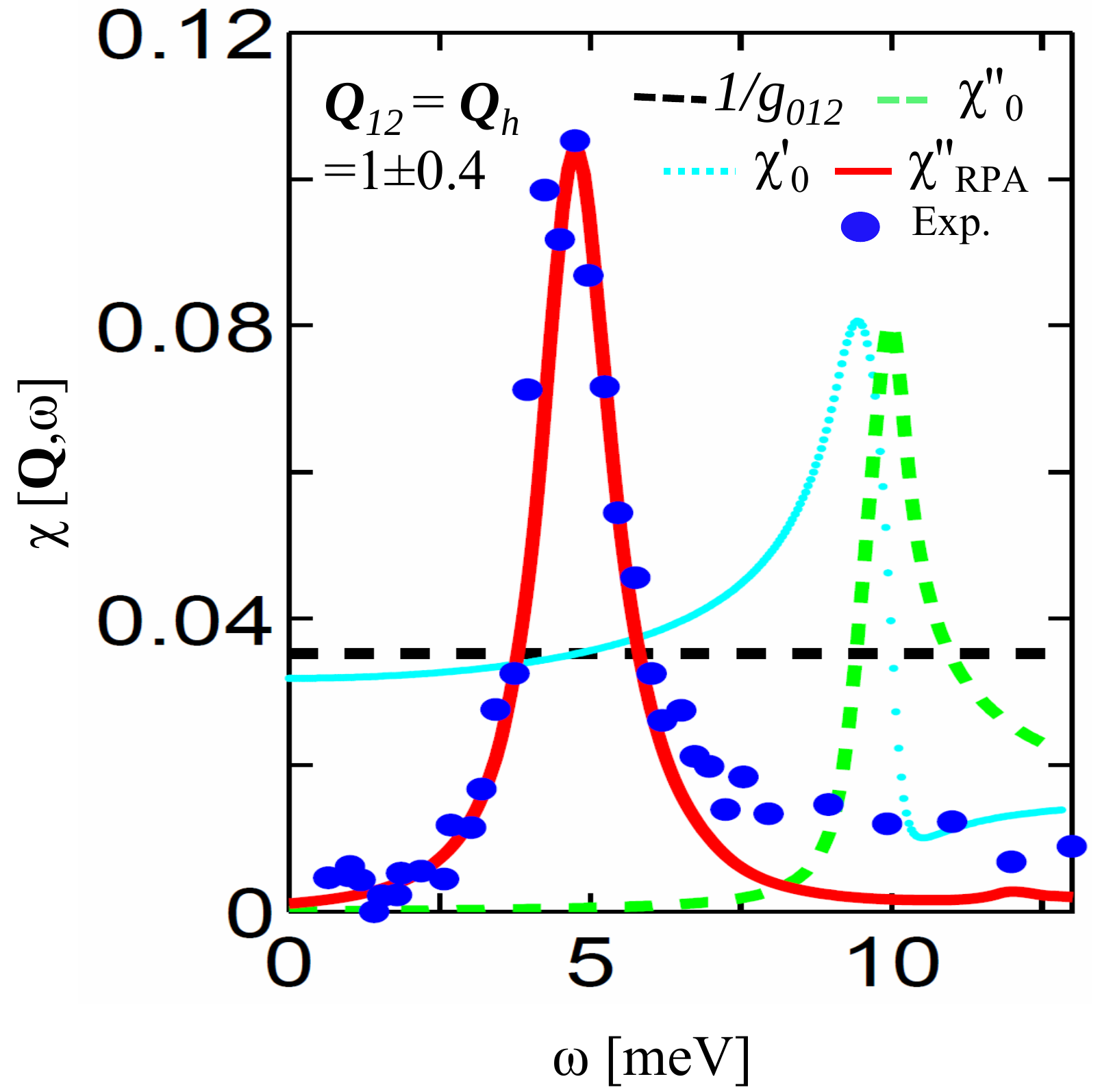}}}
\caption{The real and imaginary part of the bare susceptibility calculated from Eqs.~\ref{chi1}-\ref{chi0} in the hidden order state. The red line gives the INS mode obtained using RPA calculation at $g_{12}=28.4$~meV whose inverse is shown by horizontal dashed line. The RPA value has a huge peak (resonance) at $\omega_Q=$4.7~meV when $\chi^{\prime}({\bm Q}_h,\omega_{Q})=1/g_{12}$, and we have rescaled its intensity arbitrarily to fit it into the same figure with the bare values for comparison. The obtained resonance is in good agreement with experiment shown by symbols.\cite{INS_JPSC}}
\label{sfig2}
\end{figure}

\subsection{Neutron mode at the Hidden order state}
Eqs.~\ref{chi1} and \ref{chi0}, imply that the hidden order transition accompanies an inelastic neutron scattering mode with enhanced intensity at $Q_{h}$ whose energy scale is approximately given by $\chi^{\prime\prime}({\bm Q}_{h},\omega)\sim C\delta(\omega-E_{\bm k}^1-E_{{\bm k}+{\bm Q}_h}^2)\approx C\delta(\omega-|\Delta({\bm k})|-|\Delta({\bm k}+{\bm Q}_h)|)$. $C$ is the prefactor which has to be evaluated rigorously, but it does not contribute to the peak position in bare $\chi^{\prime\prime}$. Here we have substituted the condition that non-interacting bands are nested on the FS at $Q_{h}$ such as $E_{\bm k}^{\nu}=|\Delta(\bm{k})|\approx$5~meV. The RPA correction shifts the resonance energy which depends on the value of coupling constant. At $g_{12}=28.4$~meV, the $1/g_{12}$ line cuts twice to $\chi^{\prime}({\bm Q}_h,\omega)$, however, the resonance is stronger at the energy where the broadening function coming from $\chi^{\prime\prime}$ is weaker. Therefore, the strong intensity or a resonance occurs at $\omega_{Q}\sim4.7$~meV, see Fig.~\ref{sfig2}. INS measurements have observed this resonance at the incommensurate nesting vector $Q=(1.4,0,0)$.\cite{INS_JPSC,INS,INS_mydosh} However, one have to be careful to directly compare our result with this data. Because, in the present case, we expect a mode which does not break time-reversal symmetry. Therefore, in order to observe this mode, one requires to perform a polarized INS measurement.

On the same reasoning, INS should see more resonances at $Q_{11}$ and $Q_{22}$ even in the non-polarized condition.  However, the energy scale and intensity of those modes will depend on the location of the phase diagram of URu$_2$Si$_2$ where the experiment is performed. As we argued earlier, $Q_{11}$ is most likely responsible for the LMAF phase, therefore, we can expect a mode at twice of the corresponding gap opening. The peak in the INS spectra at the commensurate vector is also observed at a much lower in energy, however, the peak is much broader than the resonance peak observed at the incommensurate vector.\cite{INS_JPSC,INS_mydosh}

\begin{figure}[top]
\hspace{-0cm}
\rotatebox{0}{\scalebox{.6}{\includegraphics{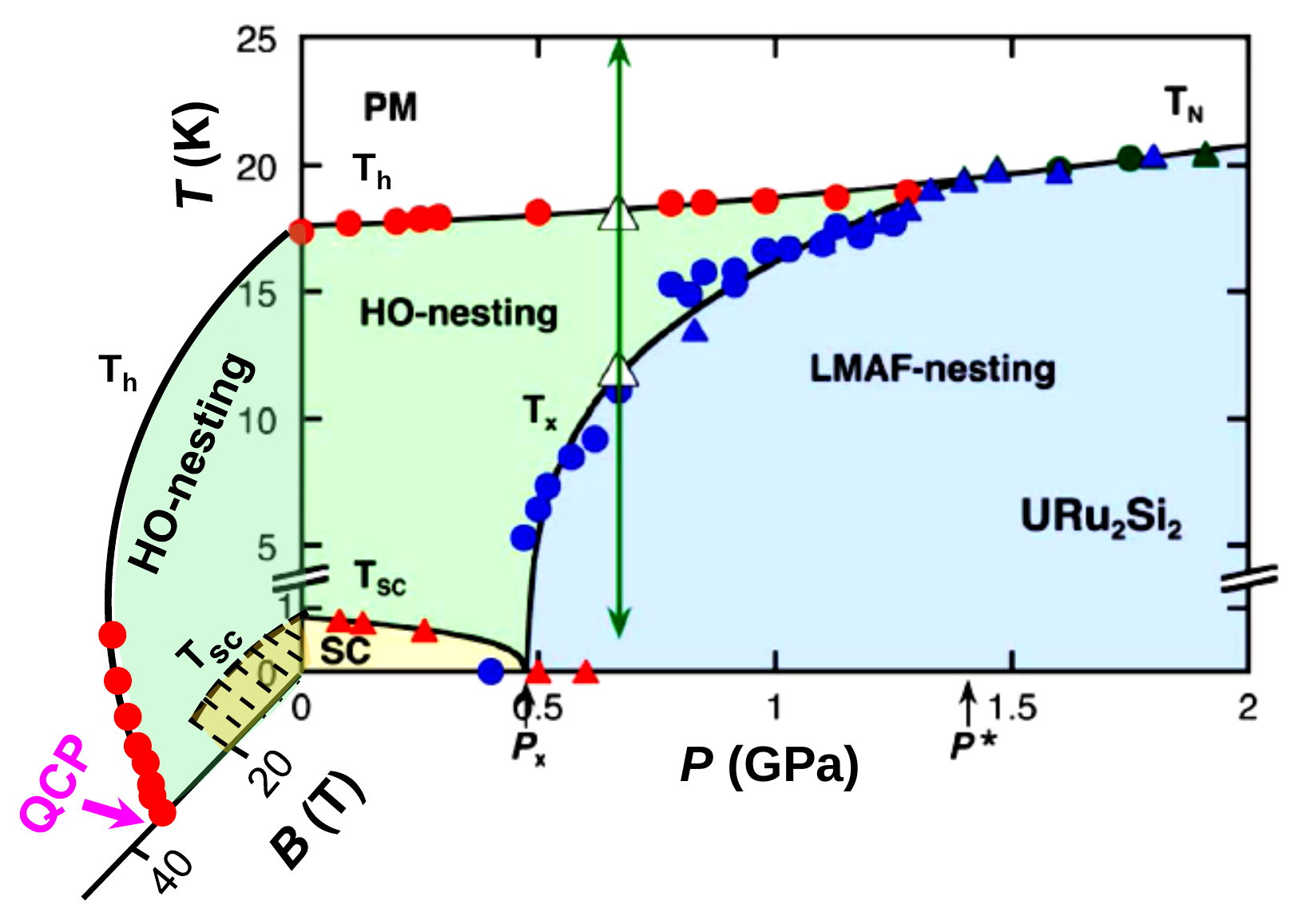}}}
\caption{ Proposed phase diagram of URu$_2$Si$_2$ along the magnetic field (${\bm B}$) direction. The data along pressure ($P)-T$ plane is taken from Ref.~\onlinecite{phase_diagram}. The symbols along $B-T$ axes are extracted from Ref.~\onlinecite{HarrisonQCP}. A QCP at $T=0$ along the field directions is expected from our theory, and also observed in experiment.\cite{HarrisonQCP} Deducing the phase diagram for SC and other possible phases that may arise above the QCP is beyond the scope of the present calculation.}
\label{sfig3}
\end{figure}

\section{Quantum Critical Point}
As mentioned in the main text, any $\mathcal{TR}$ breaking perturbation such as magnetic field will destroy the $\mathcal{TR}$ invariant HO phase. As the HO state incipiently is a spontaneously broken symmetry, it will exhibit a second order phase transition along the field axis. In what follows we expect to obtain a quantum critical point (QCP) at $T=0$, as extensively proposed to be associated with any second order phase transition,\cite{sachdev} along the magnetic field axis.  We find the critical field to be $B\approx$38~T, whereas the experimental data\cite{HarrisonQCP} indicates that the QCP resides around $B\sim34$ T. The present model can not deduce the phase diagram for the superconducting (SC) state, possibly intervening the HO state, or any other phases that may arise above the QCP.\cite{HarrisonQCP} More experimental data and theoretical modelling are required to understand the details of this regime of the phase diagram.

\end{document}